\documentclass[12pt]{article}
\usepackage{url,cite}
\usepackage{pifont}
\usepackage{times} 
\usepackage{amssymb,amsbsy}
\usepackage{amsmath,amscd}
\usepackage{graphics}
\usepackage{pstricks,pst-plot}
\renewcommand{\omega}{\varpi}
\newcommand\email[2]{#1@#2}

\def\a{\boldsymbol{a}}
\def\b{\boldsymbol{b}}
\def\k{\boldsymbol{k}}
\def\v{\boldsymbol{v}}
\def\x{\boldsymbol{x}}
\def\A{\mathcal{A}}
\def\C{{\mathbf{C}}}
\def\R{{\mathbf{R}}}
\def\P{{\mathbf{P}}}
\def\Z{{\mathbf{Z}}}
\def\G{{\Gamma}}

\def\eq#1{(\ref{#1})}

\renewcommand\hat{\widehat}
\renewcommand\tilde{\widetilde}

\begin{document}
\title{An Algebraic Geometry Method for Calculating DOS for 
2D tight binding models}
\author{ 
Koushik Ray \thanks{\email{koushik}{iacs.res.in}} \\
\small Department of Theoretical Physics,
\small  Indian Association for the Cultivation of Science \\
\small  Calcutta 700~032. India.
\and 
Siddhartha Sen  
\thanks{\email{siddhartha.sen}{tcd.ie}, \email{sen1941}{gmail.com}{} }\\ 
\small CRANN, Trinity College Dublin, Dublin -- 2, Ireland
\small \&\\
\small  R.K. Mission Vivekananda University,
\small Belur-711202, West Bengal, India.
}
\maketitle
\vfil
\begin{abstract}
\noindent 
An algebraic geometry method is used to calculate the moments of the 
electron density of states as a function of the energy  for
lattices in the tight binding approximation.
Interpreting the moments as the Mellin transform of  the density
allows writing down a formula for the density
as an inverse Mellin transform.  The method is illustrated
by working out the density function for the two-dimensional square and
honeycomb lattices.
\end{abstract}
\thispagestyle{empty}
\clearpage
\noindent 
The tight binding model is a widely used scheme for studying electronic 
band structure of solids \cite{ecnmu}. The model is defined by a 
Hamiltonian quadratic in the electron creation and destruction operators 
indexed by a set of points in the $D$-dimensional 
Euclidean space $\R^D$, called sites. 
The sites form a lattice $\Lambda$, taken to model a crystal. 
The physical picture underpinning the model supposes that the electrons
are tightly bound to a site but may hop from
a given site to its neighbouring ones, which, for the purpose of the
present discussion, are restricted to the nearest neighbours only, with 
proximity defined with respect to
distances measured along lattice paths. 
Thus each physical system is defined by its specific lattice 
description. The translation symmetry of the lattice permits
restricting the quasi-momenta $\boldsymbol k$, 
that is the variables on the reciprocal
lattice $\widetilde{\Lambda}$, 
dual to $\Lambda$, to a closed subset of the dual
$\R^D$. The convex hull of this closed
subset is called the Brillouin zone. 
The eigenvalues of the 
tight binding Hamiltonian are invariant functions defined on the 
Brillouin zone. 

We consider a related variant of the tight binding approximation
wherein the energy eigenvalues of electrons are those of
a discrete Laplacian associated with the lattice \cite{stin1}.  
The discrete Laplacian is defined on
complex-valued functions $f$ on $\R^D$ as
\begin{equation*} 
\nabla f(\v)=\sum\limits_{\a,\b\;\in\mathcal{A}\subset\Lambda}
c_{\a}c_{\b}f(\v+\a-\b),
\end{equation*} 
where the set $\mathcal{A}$ generates the lattice
$\Lambda$. The parametres $c$ are taken to be unity on every site.
The eigenvalues of the Laplacian for
this variant are the square of the energy eigenvalues obtained from the
usual tight binding model.

Given a lattice $\Lambda$ 
in $\R^D$ and the energy eigenvalues of the single electron states,
its associated Green's function, often referred to as the lattice Green's
function, can be evaluated and has found
diverse applications \cite{morita,berciu,gutt,kouts}. 
The electronic density of states (DOS) as well as a host of other 
physical quantities of the crystalline solid can be obtained
from the Green's function. For example, the DOS of the system
can be determined from the imaginary part of the Green's function $G$
as $\rho(\epsilon)=-(1/\pi)\lim_{\eta\rightarrow 0^+}
\text{Im}\, G(\epsilon+i\eta)$, $\epsilon$ denoting the energy eigenvalue. 
This follows from the definition of the density of states 
as a sum over delta functions $\delta(E-E_n)$ 
over energies, where $E_n$ is an electron energy eigenvalue.
Various techniques have been developed to determine the density of states 
as it contains important physical information, such as electron conductivity
in solids \cite{gaspard,trias,pias}. 

We use standard methods of algebraic geometry
to determine the moments of the
density of states for the tight binding model in two
dimensions. There is a certain naturality in this formulation.
First, the Laplacian is a natural operator and second, 
periodic functions of two variables, 
like the single electron eigenvalues obtained here,
is a means to define a well-studied object in algebraic geometry, namely
a complex algebraic surface, also called an elliptic curve. 
Thus all two-dimensional lattice systems
with energy eigenvalues periodic in both directions
represent elliptic curves. Let us mention that 
although we restrict to two-dimensional models only, the
technique used here generalises to higher dimensions. 

The density of states in this 
approach can be written solely in terms of the combinatorial data of the
lattice, without requiring the knowledge of electronic  
wave functions and sum over delta functions. There exist 
algebraic geometry methods for studying
elliptic curves using differential equations, 
known as Picard-Fuchs equations. The solutions to  these
equations provide an alternative way to describe an
elliptic curve. From this algebraic geometry
insight the electronic density of states can also be related
to these solutions of the Picard-Fuchs equation which,
in our case, is a single second order differential equation \cite{morrison}.
Circuits around the singular points of the Picard-Fuchs equation 
are related to the (co)homological properties of the curve.
Indeed, the derivation of the Picard-Fuchs equation follows from
these topological properties. A surface, such as the Brillouin zone,
which is doubly periodic, is topologically a torus, with two
linearly independent closed one-forms that are not exact. 
Let us recall that a closed one-form on a space
is one that vanishes when operated on by the differential
operator $d$. It can be written locally, though not necessarily globally,
as $df$ where $f$ is a function on the space. An
exact one-form is one that can be globally written as $df$. Such a form
becomes identically zero when operated on by the operator $d$.
The dimension of the first cohomology group of the space is the number of 
linearly independent closed but not exact one-forms.
It is a topological invariant. 
In the two-dimensional examples that we discuss, 
there are two linearly independent closed but not exact one-forms present.  
Thus, if we start with an arbitrary local expression for a family of
one-forms on the surface and differentiate with respect to the family 
parametre $z$, then every differentiation produces a new one-form.
Thus, the first and second derivatives along with the original one make three
one-forms. If all of these one-forms are further
constructed to be closed then we know that there must be a
linear relationship between these three since the first cohomology group has
dimension two. This linear relationship is the Picard-Fuchs equation. 
The procedure of constructing the one-forms on algebraic surfaces
through differentiation with
respect to the family parametre and discarding exact one-forms at each step
has been used earlier in various contexts 
\cite{morrison,schnell,isidro}.
 
We illustrate this approach in two examples, namely,
the two-dimensional tight binding model for the square and honeycomb lattices.
The honeycomb case represents graphene which is a system of
considerable current interest.
In these two cases we show the two steps used to  determine 
the electron energy density of states. First, algebraic geometry
is used to determine the energy moments
of the density of states which are interpreted as
Mellin transforms. Next, we use the powerful 
techniques of inverting Mellin transforms to determine an analytic 
expression for the density of states.
Indeed, the advantage of the present approach
lies in obtaining the density of states as an inverse Mellin transform,
simplifying numerical evaluations for any value of the energy. 

Let us start by briefly discussing the general combinatorial set up to 
fix notation. We shall also identify the physical quantities, in particular,
the density of states, in terms of the 
combinatorial data. We restrict the discussion
to two-dimensional cases. Generalisation 
to higher dimensions may be considered following known results \cite{stin1}. 
The model we consider is described by a finite subset $\mathcal{A}$ of
$\Z^2$. The lattice $\Lambda$
is then obtained by taking the $\Z$-span of the difference of points in
$\mathcal{A}$, that is 
\begin{equation} 
\Lambda=\Z\text{-span}\{\a-\b|\ \a,\b\;\in\mathcal{A}\}.
\end{equation} 
In other words, the set $\A$ is obtained as marking one of the lattice points
of the model as the origin and collecting the points connected to it by a
single path in the lattice. For example,
the set $\A = \{ (-1,0),(1,0),(0,-1),(0,1) \}$
for the square lattice, while $\A =\{(1,0),(0,1),(-1,-1) \}$ for the
honeycomb lattice in two dimensions. The 
lattice constant is taken to be unity throughout.

On the points of
$\mathcal{A}$ we consider the distribution given by a sum of Dirac deltas as
\begin{equation}
\mathcal{D}=\sum\limits_{\a\;\in\mathcal{A}}\delta_{\a}. 
\end{equation} 
Being supported solely on the lattice points, this embodies a crystal in
the tight binding approximation. 
The delta functions may, in principle, have different weights at different
points, but we shall not consider that here.
The Fourier transform of $\mathcal{D}$ is given by 
\begin{equation} 
\hat{\mathcal D}(\k)=\sum\limits_{\a\;\in\A} e^{-2\pi i\k\cdot\a},
\end{equation} 
where the quasi-momenta $\k=(k_1,k_2)$ are valued in the reciprocal lattice
\begin{gather}
\tilde{\Lambda}=\{\k\;\in\R^2|\k\cdot (\a-\b)\;\in\Z,\forall\a,\b\;\in\mathcal{A}\}
\end{gather} 
dual to $\Lambda$.
The eigenvalues of the discrete 
Laplacian $\nabla$ based on $\Lambda$ are then written
in terms of the quasi-momenta as the dispersion relation
\begin{equation}
\label{eq:disp}
E(\k)^2:=|\widehat{\mathcal{D}}(\k)|^2 
= \sum\limits_{\a,\b\;\in\mathcal{A}} \cos 
2\pi\k\cdot (\a-\b).
\end{equation} 
The energy 
$E$ is periodic with period lattice $\tilde{\Lambda}$ thus descending to a 
function on the Brillouin zone $U^{\Lambda}\sim\R^2/\tilde{\Lambda}$, 
which has the topology of a torus. 
Let us introduce complex variables $x$, $y$ and define a Laurent polynomial
\cite{stin1}
\begin{equation}
W(x,y) = \sum\limits_{\a,\b\,\in\A}{\x}^{\a-\b},
\end{equation} 
associated to the set $\A$, 
satisfying $|\widehat{\mathcal{D}}(\k)|^2
=W(e^{2\pi ik_1},e^{2\pi ik_2})$, where
$\x = (x_1,x_2)=(x,y)$ and ${\x}^{\boldsymbol\lambda} =
x_1^{\lambda_1}x_2^{\lambda_2}$, for $\boldsymbol\lambda\in\Lambda$. 
The number of states, denoted $V(\epsilon)$, is given by the 
normalized volume of the Brillouin zone such that 
$|\widehat{\mathcal{D}}(\k)|^2\leq\epsilon$. Let us remark that, as
mentioned before, by equation \eq{eq:disp}, the parametre $\epsilon$ is
the square of the energy obtained from an usual tight binding model. 
The Hilbert transform of the differential $dV$ is defined as the integral 
of the resolvent $1/(z-\epsilon)$ with respect to the measure 
defined by $dV$ over the real line as
\begin{align}
H(z) &= \int_{\R}\frac{dV(\epsilon)}{z-\epsilon} \label{hlb:def}\\
&=\frac{1}{(2\pi i)^2}
\int\limits_{|x|=|y|=1}\frac{1}{z-W(x,y)}\frac{dx}{x}\frac{dy}{y}.
\label{hlb:per}
\end{align}
where $z$ is a complex parametre.

The function $H(z)$ in \eq{hlb:per}
is the period of a differential one-form along a
one-cycle on the hypersurface given by $z=W(x,y)$ in $(\C^{\star})^2$. 
It is obtained as a solution to a Picard-Fuchs equation in the 
form of a Laurent series in the complex variable $z$ which, 
according to \eq{hlb:def}, is given in terms of  moments $a_n$ as
\begin{equation}
\label{hlb:mom}
H(z)=\sum\limits_{n=0}^{\infty}a_nz^{-1-n}.
\end{equation}
The moments can be calculated by either substituting \eq{hlb:mom} in the 
Picard-Fuchs equation or by using the residue theorem in \eq{hlb:per} as
\begin{equation*}
a_n = \text{constant term of the Laurent polynomial} ~W(x,y)^n.
\end{equation*}
From  the moments one can calculate the lattice Green's function 
and hence the density of states \cite{trias,gaspard}.
Although these methods  do not yield explicit formulas, they 
lead to systematic approximation schemes that can be numerically 
implemented in an efficient manner. We shall consider
a different way to obtain the density of states from $H(z)$ which yields
explicit formulas. The idea
is to expand \eq{hlb:def} in a geometric series in $\epsilon/z$ 
as 
\begin{align}
H(z) &= \frac{1}{z}\sum\limits_{n=0}^{\infty}
\int_{\R} \frac{dV}{d\epsilon} (\frac{\epsilon}{z})^n d\epsilon\\
& = \sum\limits_{n=0}^{\infty} z^{-1-n}  \int_{\R} 
\rho(\epsilon) \epsilon^n d\epsilon, 
\end{align}
where we defined the density of states $\rho(\epsilon)=dV/d\epsilon$.
Comparing with \eq{hlb:mom} we conclude, 
\begin{equation}
a_n  =  \int_{\R} \rho(\epsilon)\epsilon^n d\epsilon.
\end{equation}
We now make our simple but important observation, namely, the moments $a_n$
of the density of states $\rho(\epsilon)$ can be interpreted as 
the Mellin transform of $\epsilon \rho(\epsilon)$ 
if we replace the integers $n$ by a complex
variable $s$. An immediate consequence of this remark is, 
as emphasised before, that an expression for 
the density of states can be easily written down 
as the inverse Mellin transform of $a_n=a(s)$. We have the formula 
\begin{equation}
\label{iMel}
\rho(\epsilon) = \frac{1}{2\pi i}\int\limits_{c_0-i\infty}^{c_0+i\infty}
\epsilon^{-1-s}a(s) ds,
\end{equation} 
where the line integral is evaluated along a vertical line in the complex plane
and $c_0$ is an appropriate real constant.
This approach thus gives an explicit formula for the density of states
in terms of a function determined by the methods of algebraic geometry. 
Moreover,
it allows us to calculate the density of states for any value of $\epsilon$,
large or small, by choosing appropriate contours in the $s$-plane. 
In order to use this method we need to be able to
replace the discrete set $a_n$ by a function $a(s)$ of a complex variable $s$.
For the cases that we study there is a natural way of doing this.
We shall now consider two examples. 
\subsubsection*{Example~1: Square lattice}          
For the square lattice the set of generating points in $\Z^2$ is
$\mathcal{A}=\{(-1,0),(1,0),(0,-1),(0,1)\}$. This corresponds to the polynomial 
\begin{equation}
W = (x+1/x+y+1/y)^2 
\end{equation} 
in the coordinate ring $\C[x,x^{-1},y,y^{-1}]$.
Then the dispersion relation is obtained to be
\begin{equation}
E(\k)^2=4+2\cos 2\pi k_1 + 2\cos 2\pi k_2.
\end{equation}
We shall evaluate the resolvent $H(z)$ defined in \eq{hlb:per}. 
Writing the complex variables $x,y$ in terms of the homogeneous 
coordinates of a two-dimensional complex projective space $\P^2$
as $x=v_1/v_0$ and $y=v_2/v_0$, $v_0\neq 1$, we rewrite $H$ as 
\begin{equation}
\begin{split}
H(z) &=\int\frac{v_0v_1v_2 \Omega} {z (v_0v_1 v_2)^2 -
(v_1+v_2)^2(v_1v_2+v_0^2)^2},\\
&= \frac{1}{2z}\left(
\int\frac{\Omega}{ (v_0v_1 v_2) - t (v_1+v_2)(v_1v_2+v_0^2)}
+\int\frac{\Omega}{ (v_0v_1 v_2) + t (v_1+v_2)(v_1v_2+v_0^2)}
\right),
\end{split}
\end{equation} 
where $\Omega = v_0 dv_1\wedge dv_2-v_1dv_0\wedge dv_2+v_2 dv_0\wedge dv_1$ 
is the canonical $2$-form on $\P^2$ and we defined $t=1/\sqrt{z}$.
The Picard-Fuchs equations of both the varieties 
\begin{equation} 
t (v_1+v_2)(v_1v_2+v_0^2)\pm  v_0v_1v_2=0,  
\end{equation} 
are the same, namely,
\begin{equation}
\frac{d^2\omega}{dt^2}+\frac{1-48t^2}{t-16t^3}\frac{d\omega}{dt}
+\frac{16\omega }{16t^2-1}=0. 
\end{equation} 
Thus, series solutions with only the terms with even powers of $t$ survive. 
This Fuchsian equation has two solutions which can be obtained as series in
$t$ by the Frobenius' method. The two solutions are
\begin{gather}
\omega_1(t) = {}_2F_1(1/2,1/2;1;16t^2),\\
\omega_2(t) = (\log t+2\log 2) {}_2F_1(1/2,1/2;1;16t^2) 
+ \frac{1}{2\pi} \sum\limits_{n=0}^{\infty}
\frac{d}{d\alpha}\left[
\frac{\Gamma(\alpha+n+1/2)^2}{\Gamma(\alpha+m+1)^2}
\right]_{\alpha=0} (16t^2)^n.
\end{gather}
Here ${}_2F_1(a_1,a_2;b_1;x)$ represents a hypergeometric function defined
by the series
\begin{equation} 
{}_2F_1(a_1,a_2;b_1;x)=\sum\limits_{n=0}^{\infty}
\frac{(a_1)_n(a_2)_n}{(b_1)_n\ n!}x^n,
\end{equation} 
where $(a)_n=a(a+1)...(a+n-1)$ is the Pochhammer symbol.
The two integrals in  $H$ are then linear combinations of the two solutions,
namely,
\begin{equation} 
\begin{split}
H(z) &= \frac{c_1}{2z}{}_2F_1(1/2,1/2;1;\frac{16}{z}) \\
&+ \frac{c_2}{2z} 
(\log z-4\log 2) {}_2F_1(1/2,1/2;1;\frac{16}{z})
- \frac{c_2}{4\pi z} \sum\limits_{n=0}^{\infty}
\frac{d}{d\alpha}\left[
\frac{\Gamma(\alpha+n+1/2)^2}{\Gamma(\alpha+n+1)^2}
\right]_{\alpha=0} (\frac{16}{z})^n,
\end{split}
\end{equation} 
where $c_1$ and $c_2$ are arbitrary constants. Instead of trying to determine
the constants from boundary conditions, we shall recourse to the calculation
of moments to determine the density of states.  
This entails direct evaluation of the integral \eq{hlb:per} using residues.
Since the constant term in the expansion of $W^n$ is $\binom{2n}{n}^2$,
we have 
\begin{equation} 
\begin{split}
a_n &= \binom{2n}{n}^2,  \\
&= \frac{\G(2n+1)^2}{\G(1+n)^4}\\
&= \frac{(2n)^2 \G(2n)^2 }{n^2\G(n)^2\G(1+n)^2}\\
&= \frac{1}{\pi}\frac{\G(n+1/2)^2}{\G(1+n)^2}{16}^n,
\label{aen:sq}
\end{split}
\end{equation} 
where we used the duplication formula
$\pi^{1/2}\G(2x)=2^{2x-1}\G(x)\G(x+1/2)$ in the last step.
By \eq{hlb:mom} this gives the resolvent $H$ as
\begin{equation}
\begin{split}
H(z) &=  \sum\limits_{n=0}^{\infty} a_n z^{-1-n}\\
&= \frac{1}{2z}{}_2F_1(1/2,1/2;1;\frac{16}{z}).
\end{split}
\end{equation} 
By \eq{iMel}, the density of states is then obtained as the inverse Mellin
transform 
\begin{equation}
\rho(\epsilon) = \frac{1}{2\pi i}\frac{1}{\pi}
\int\limits_{c_0-i\infty}^{c_0+i\infty}
\frac{\G(s+1/2)^2}{\G(1+s)^2}{16}^s\epsilon^{-1-s},
\end{equation} 
where the integrand is derived from \eq{aen:sq} by substituting $s$ for $n$.
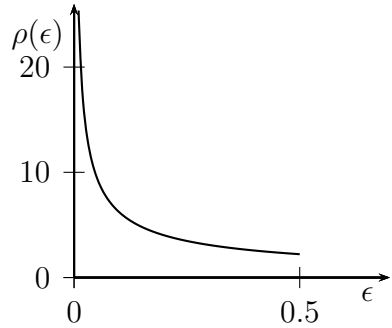
\begin{figure}[h]
\begin{center}
\begin{pspicture}(-1,-.7)(4,4)
\psset{xunit=6cm,yunit=1.4mm}
\savedata{\mydata}[
{{0.01, 25.3791}, {0.02, 16.7222}, {0.03, 13.0698}, {0.04, 10.9604}, {0.05, 
    9.55475}, {0.06, 8.53704}, {0.07, 7.75889}, {0.08, 7.14047}, {0.09, 
    6.6346}, {0.1, 6.21145}, {0.11, 5.85111}, {0.12, 5.53975}, {0.13, 
    5.26743}, {0.14, 5.02678}, {0.15, 4.81226}, {0.16, 4.61955}, {0.17, 
    4.44528}, {0.18, 4.28675}, {0.19, 4.14179}, {0.2, 4.00862}, {0.21, 
    3.88575}, {0.22, 3.77196}, {0.23, 3.66621}, {0.24, 3.56762}, {0.25, 
    3.47545}, {0.26, 3.38903}, {0.27, 3.30782}, {0.28, 3.23132}, {0.29, 
    3.15911}, {0.3, 3.09081}, {0.31, 3.0261}, {0.32, 2.96467}, {0.33, 
    2.90627}, {0.34, 2.85066}, {0.35, 2.79764}, {0.36, 2.74701}, {0.37, 
    2.69861}, {0.38, 2.65228}, {0.39, 2.60788}, {0.4, 2.56529}, {0.41, 
    2.52439}, {0.42, 2.48507}, {0.43, 2.44724}, {0.44, 2.41081}, {0.45, 
    2.3757}, {0.46, 2.34184}, {0.47, 2.30914}, {0.48, 2.27756}, {0.49, 
    2.24702}, {0.5, 2.21748}}
]
\psaxes[Dx=.5,Dy=10]{->}(0,0)(.7,26)
\rput(.65,-1.6){$\epsilon$}
\rput(-.08,23){$\rho(\epsilon)$}
\dataplot[plotstyle=curve,showpoints=false]{\mydata}
\end{pspicture}
\caption{Behaviour of density of states $\rho(\epsilon)$ near $\epsilon=0$ for
the square lattice}
\label{sq:dos}
\end{center}
\end{figure}
Choosing $c_0=0$ and closing the contour with a semicircular arc on the left
so as to obtain an expression valid near $\epsilon=0$, we  get
\begin{equation}
\rho(\epsilon) = \frac{4\log
2-\log\epsilon}{4\sqrt{\epsilon}}{}_2F_1(1/2,1/2;1;\epsilon/16)+
\frac{\pi}{4\sqrt{\epsilon}}\sum\limits_{n=0}^{\infty}
\frac{d}{ds}\left[\frac{1}{\G(1+s)^2\G(1/2-s)^2} \right]_{s=-1/2-n}
\left(\frac{\epsilon}{16}\right)^{n},
\end{equation} 
whose leading behaviour near $\epsilon=0$ is shown in Figure~\ref{sq:dos}.
Let us point out that there is a lower limit to the admissible range of
energy $\epsilon$ determined by 
the sample geometry. The density of states plotted integrated 
over the whole range of energy does not depend on this physical cut off since
the integral is finite even without a cutoff. The same holds good
for the honeycomb lattice as well to which we now turn as our next example.
\subsubsection*{Example~2: honeycomb lattice}
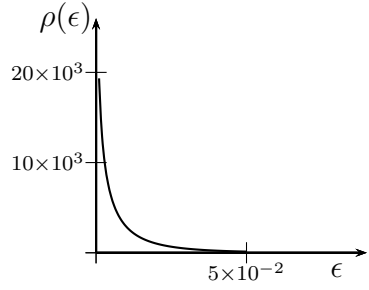
\begin{figure}[h]
\begin{center}
\begin{pspicture}(-1,-.7)(4,4)
\psset{xunit=40cm,yunit=1.2mm}
\savedata{\mydata}[
{{0.001, 19.337}, {0.002, 12.9066}, {0.003, 9.66994}, {0.004, 
    7.64983}, {0.005, 6.25349}, {0.006, 5.22758}, {0.007, 4.44229}, {0.008, 
    3.82309}, {0.009, 3.32371}, {0.01, 2.91374}, {0.011, 2.57228}, {0.012, 
    2.28446}, {0.013, 2.03938}, {0.014, 1.82889}, {0.015, 1.64674}, {0.016, 
    1.48807}, {0.017, 1.34905}, {0.018, 1.22661}, {0.019, 1.11827}, {0.02, 
    1.02199}, {0.021, 0.936111}, {0.022, 0.859234}, {0.023, 0.790195},
{0.024,
     0.72801}, {0.025, 0.671846}, {0.026, 0.62099}, {0.027, 0.574832},
{0.028,
     0.532844}, {0.029, 0.494571}, {0.03, 0.459616}, {0.031, 
    0.427634}, {0.032, 0.398321}, {0.033, 0.371411}, {0.034, 
    0.346669}, {0.035, 0.323886}, {0.036, 0.30288}, {0.037, 0.283485},
{0.038,
     0.265556}, {0.039, 0.248962}, {0.04, 0.233586}, {0.041, 
    0.219324}, {0.042, 0.20608}, {0.043, 0.19377}, {0.044, 0.182318}, {0.045, 
    0.171653}, {0.046, 0.161712}, {0.047, 0.152439}, {0.048, 
    0.143781}, {0.049, 0.135692}, {0.05, 0.128128}}
]
\psaxes[Dx=.05,Dy=10,labels=none]{->}(0,0)(.09,26)
\rput(-.016,10){$\scriptstyle 10\times 10^3$}
\rput(-.016,20){$\scriptstyle 20\times 10^3$}
\rput(.05,-2.2){$\scriptstyle 5\times 10^{-2}$}
\rput(.08,-2){$\epsilon$}
\rput(-.01,26){$\rho(\epsilon)$}
\dataplot[plotstyle=curve,showpoints=false]{\mydata}
\end{pspicture}
\caption{Behaviour of density of states $\rho(\epsilon)$ near $\epsilon=0$ for
the honeycomb  lattice}
\label{hon:dos}
\end{center}
\end{figure}

For the honeycomb lattice we have $\A =\{(1,0),(0,1),(-1,-1)\}$, leading to
the  Laurent polynomial 
\begin{equation}
W(x,y)=(x+y+\frac{1}{xy})(\frac{1}{x}+\frac{1}{y}+xy)
\end{equation}
in $\C[x,x^{-1},y,y^{-1}]$.
The dispersion relation is 
\begin{equation}
E(\k)^2= 3 + \cos\bigl(2\pi(k_1-k_2)\bigr) + \cos\bigl(2\pi (2k_1+k_2)\bigr) 
+ \cos\bigl(2\pi (k_1+2k_2)\bigr),
\end{equation}
which upon a change of basis of the reciprocal  lattice
\begin{equation}
k_1 = (\sqrt{3}\kappa_x + 3\kappa_y)/6,\quad
k_2 = (\sqrt{3}\kappa_x - 3\kappa_y)/6,
\end{equation}
yields the more usual form \cite{wallace}
\begin{equation}
E(\boldsymbol\kappa)^2 = 
1+ 4\cos^2\pi\kappa_y + 4\cos\pi\kappa_y\cos\pi\sqrt{3}\kappa_x.
\end{equation}
Let us define the homogeneous coordinates of a $\P^2$, namely $[v_0:v_1:v_2]$, 
related to $x,y$ by
\begin{equation}
x^2y=v_1/v_0,\quad xy^2=v_0/v_2.
\end{equation}
Substituting these in \eq{hlb:per} we obtain the resolvent 
\begin{equation}
H(z)=\int\frac{\Omega}{z v_0v_1v_2-(v_0+v_1+v_2)(v_0v_1+v_1v_2+v_2v_0)},
\end{equation}
solving the Picard-Fuchs equation 
\begin{equation}
\label{pf:hon}
\frac{d^2\omega}{dz^2}+\frac{9-20z+3z^2}{z(9-10z+z^2)}
\frac{d\omega}{dz} + \frac{(z-3)\omega}{z(9-10z+z^2)}=0.
\end{equation}
Again, instead of writing down all the solutions of this equation, 
it suffices for our purposes to consider the moments.
The constant term in the expansion of $W$ gives the  moments \cite{stin1}
\begin{align} 
a_n &=\sum\limits_{j=0}^n\binom{n}{j}^2\binom{2j}{j} \nonumber \\
&=\sum\limits_{j=0}^n\frac{\G(1+n)^2\G(2j+1)}{\G(1+n-j)^2\G(1+j)^4} 
\nonumber \\
&=\frac{1}{\sqrt{\pi}}
\sum\limits_{j=0}^n\frac{\G(1+n)^2\G(j+1/2)}{\G(1+n-j)^2\G(1+j)^3} 4^{j} 
\displaybreak[0]\nonumber \\
&=\frac{1}{\sqrt{\pi}}\sum\limits_{j=0}^{\infty}
\frac{\G(1+n)^2\G(j+1/2)}{\G(1+n-j)^2\G(1+j)^3} 4^{j}, \label{aen:hon} \\
&= {}_3F_2({1/2,-n,-n};{1,1};4),
\end{align} 
where ${}_3F_2(a_1,a_2,a_3;b_1,b_2;x)$ is the generalised
hypergeometric function defined by the series
\begin{equation} 
{}_3F_2(a_1,a_2,a_3;b_1,b_2;x)=
\sum\limits_{n=0}^{\infty}(\frac{(a_1)_n(a_2)_n(a_3)_n}{(b_1)_n(b_2)_n\
n!}x^n.
\end{equation} 
The duplication formula has been used 
in deriving the expression \eq{aen:hon} and 
the sum has been extended to all integral values of $j$ 
since $1/\G(1+n-j)$ vanishes for all $j\geq n+1$.
As before, equation \eq{pf:hon} is solved with 
\begin{equation}
H(z)=\sum\limits_{n=0}^{\infty} a_n z^{-1-n}.  
\end{equation} 
Then the density of states is expressed in terms of the inverse Mellin
transform of $a(s)$ as
\begin{equation}
\rho(\epsilon) = \frac{1}{\sqrt{\pi}} \frac{1}{2\pi i}
\sum\limits_{k=0}^{\infty}
\int\limits_{c_0-i\infty}^{c_0+i\infty}
\frac{\G(1+s)^2\G(k+1/2)}{\G(1+s-k)^2\G(1+k)^3}  4^k\epsilon^{-1-s}.
\end{equation} 
We can also write the sum over $k$ as an integral, as
\begin{equation}
\rho(\epsilon) = \frac{1}{\sqrt{\pi}} (\frac{1}{2\pi i})^2
\int\limits_{\mathcal{C}}dt
\int\limits_{c_0-i\infty}^{c_0+i\infty}ds
\frac{\G(1+s)^2\G(t+1/2)\G(-t)}{\G(1+s-t)^2\G(1+k)^2}  4^t\epsilon^{-1-s},
\end{equation} 
where the contour $\mathcal{C}$ is chosen so as to go parallel to the imaginary
axis and closing on the right to enclose integers $t=0,1,2,\cdots$ on the
$t$-plane. 
Now reversing the order of the integrations
we first work evaluate the integral over $t$ by closing the contour on
the left, $\text{Re}(t) <0$, so that we pick up contributions from the poles
of $\G(1/2+t)$ at $t=-1/2-k$, for positive integers $k$. This leads to 
\begin{equation}
\rho(\epsilon) = \frac{1}{2\sqrt{\pi}} \frac{1}{2\pi i}\sum\limits_{k=0}^{\infty}
\int\limits_{c_0-i\infty}^{c_0+i\infty}
\frac{\G(1+s)^2\G(k+1/2)}{\G(3/2+k+s)^2\G(1/2-k)^2\G(1+k)} 
4^{-k}\epsilon^{-1-s}.
\end{equation} 
In order to derive a power series in $\epsilon$, we note that
$\G(1+s)^2$ has double poles at $s=-1-n$, for positive integral $n$. Thus, 
performing the integral by closing the contour on the left we obtain
\begin{equation}
\rho(\epsilon) = \frac{1}{2\sqrt{\pi}}\sum\limits_{n=0}^{\infty}
\sum\limits_{n=0}^{\infty}
\frac{d}{ds}\left[\frac{\G(k+1/2) 4^{-k}\epsilon^{-1-s}}
{\G(3/2+k+s)^2\G(1/2-k)^2\G(1+k)\G(-s)} 
\right]_{s=-1-n},
\end{equation} 
which can be rewritten as
\begin{equation} 
\begin{split}
\rho(\epsilon)=\frac{1}{2\sqrt{\pi}}\sum\limits_{n=0}^{\infty}
\sum\limits_{k=0}^{\infty}
\frac{(1/4)^k \G(1/2+k)}{\G(1/2-k)^2\G(1+k)}
\epsilon^n     
&\left(
-\frac{1}{\G(1/2+k-n)^2\G(1+n)^2}
\log\epsilon \right.\\
&\left.+ 
\frac{d}{ds}\left[\frac{1}{\G(3/2+k+s)^2\G(-s)^2}\right]_{s=-1-n}
\right).
\end{split}
\end{equation} 
The behaviour of the density of states near $\epsilon=0$ is plotted in
Figure~\ref{hon:dos}.

To summarise, we have discussed the 
density of states of the two-dimensional nearest neighbour 
tight binding Hamiltonian from an algebraic geometry viewpoint. 
We have discussed two
examples based on the two-dimensional square and 
honeycomb lattices. The density of states is obtained as a function of
energy. The Hilbert transform of the density of states is the
resolvent that satisfies Picard-Fuchs equations of algebraic 
varieties that correspond to the lattices in a combinatorial fashion. 
Explicit expressions are given for small energies 
in terms of infinite series, involving hypergeometric functions. 
Let us note that the Brillouin zone corresponding to
each of the lattices is a topological torus. 
Thus qualitative features of the results may be understood in topological
terms. We intend to present details of these topological arguments in
a future work. A practical
advantage of this approach is that it allows evaluation of 
density of states in any domain of energy by appropriate choice of 
contours suitably in the integrals. 
The resulting infinite series obtained converge 
rather fast and may be  easily evaluated numerically. 
Finally, let us mention that we have presented our calculations 
in the context of electrons in
a crystalline medium but the results obtained are also applicable for the
density of states  of a system of phonons where a tight binding 
nearest neighbour model is appropriate \cite{weaire}.
\subsection*{Acknowledgement}
KR thanks Indra Dasgupta, Avijit Mukherjee and Krishnendu Sengupta 
for useful conversations.
SS would like to thank the Department of Theoretical
Physics, IACS where this work was done, for support.


\begin{thebibliography}{99}
\bibitem{ecnmu}
E.~N.~Economou, \emph{Green's functions in quantum physics}, Springer-Verlag,
Berlin; New York:1979.
\bibitem{stin1}
J.~Stienstra, \emph{Motives from Diffraction}, 	arXiv:math/0511485.
\bibitem{morita}
T.~Morita, \emph{Useful Procedure for Computing the Lattice Green's
Function‐Square, Tetragonal, and bcc Lattices}, 
J. Math. Phys. {\bf 12}, 1744 (1971)
\bibitem{berciu}
M.~Berciu, \emph{On computing the square lattice Green's function without any
integrations}, J. Phys. A: Math. Theor. {\bf 42}, 395207 (2009).
\bibitem{gutt}
A.~Guttmann, \emph{Lattice Green functions in all dimensions},
arXiv:1004.1435.
\bibitem{kouts}
C.~Koutschan, \emph{Lattice Green's Functions of the Higher-Dimensional
Face-Centered Cubic Lattices},  arXiv:1108.2164
\bibitem{gaspard}
J.~Gaspard, F.~Cyrot-Lackmann, \emph{Density of states from moments. 
Application to the impurity band}, Journal of Physics {\bf C6}, 3077 (1973).
\bibitem{trias}
A.~Trias, M.~Kiwi, and M.~Weissmann, 
\emph{Reconstruction of the density of states from its moments}
Phys. Rev. {\bf B28}, 1859 (1983). 
\bibitem{pias}
R.~Piasecki, \emph{Density of electron states in a rectangular lattice under
uniaxial stress}, arXiv:0804.1037.
\bibitem{morrison}
D.~Morrison, \emph{Picard-Fuchs equations and mirror maps for hypersurfaces},
Essays on Mirror Manifolds (S.-T. Yau, ed.), International Press, Hong Kong,
1992, pp. 241-264; arXiv:alg-geom/9202026.
\bibitem{schnell}
C.~Schnell, \emph{On computing Picard-Fuchs equations}, Unpublished notes at
\url{http://homepages.math.uic.edu/~cschnell/pdf/notes/picardfuchs.pdf}.
\bibitem{isidro}
J.~Isidro, A.~Mukherjee, J.~Nunes and H.~Schnitzer,
\emph{A New derivation of the Picard-Fuchs equations for effective $N=2$
superYang-Mills theories},
  Nucl.\ Phys.\  {\bf B492}, 647 (1997) [arXiv:hep-th/9609116].
\bibitem{wallace}
P.~Wallace, \emph{The Band Theory of Graphite}, 
Phys. Rev. {\bf 71}, 622 (1947). Erratum, {\it ibid} {\bf 72}, 258 (1947). 
\bibitem{weaire} D.\ Weaire \emph{Private communication to SS}.
\end{thebibliography}
\end{document}